%% file: mbl06.tex
\documentclass[prb,aps,twocolumn,reprint,floatfix,showpacs,10pt]{revtex4-1}
\usepackage{graphicx}
\usepackage{amsmath}
\usepackage[colorlinks,bookmarks=false,citecolor=blue,linkcolor=red,urlcolor=black]{hyperref}
\usepackage{umoline}

\begin{document}
\title{Impact of geometry on many-body localization }
\author{Dariusz Wiater$^{1,2}$}
\email{wiater.dariusz@gmail.com}
\author{Jakub Zakrzewski$^{1,3}$} 
 \email{jakub.zakrzewski@uj.edu.pl}

\affiliation{
\mbox{$^1$ Instytut Fizyki imienia Mariana Smoluchowskiego, Uniwersytet Jagiello{\'n}ski, ulica \L{}ojasiewicza 11, PL-30059 Krak\'ow, Poland }
\mbox{$^2$ Institute of Physics, Polish Academy of Sciences, Aleja Lotnikow 32/46, PL-02668 Warsaw, Poland}
\mbox{$^3$ Mark Kac Complex Systems Research Center,
Uniwersytet Jagiello\'nski, Krak\'ow, Poland }
}
\begin{abstract}
The impact of geometry on many-body localization is studied on simple, exemplary systems amenable to exact diagonalization treatment. The crossover between ergodic and MBL phase for uniform as well as quasi-random disorder is analyzed using statistics of energy levels. It is observed that the transition to many-body localized phase is correlated with the number of nearest coupled neighbors. The crossover from extended to localized systems is  approximately described by the so called plasma model. 
\end{abstract}
\date{\today}

\maketitle

\begin{figure}
\includegraphics[width=\columnwidth]{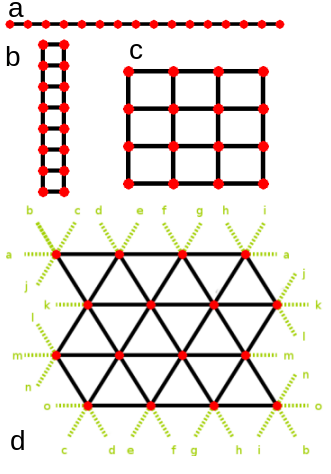}
\caption{Geometries  
considered: 
 (a) one-dimensional chain, (b) 2D ladder-like model with 8 sites in two columns, (c) a square lattice, and (d) the triangular lattice. We assume periodic boundary conditions (PBC), formulating them for a triangular lattice is nontrivial. Letters in (d) panel indicate which links are to be identified, i.e. "a" with "a", "b" with "b" etc.}
\label{Fig1}
\end{figure}

\section{Introduction}
Study  of  localization phenomena in many-body interacting systems started, arguably, in the eighties \cite{Fleishman80} but received a real start with
 Ref.~\cite{Basko06} where, using perturbative arguments, it was shown that even highly excited states may be localized in interacting many-body systems for sufficiently strong disorder.  The phenomenon was termed many-body localization (MBL) and received a lot of attention in last 10 years due to several reasons. 

Recent developments in experimental techniques allow the researchers to study isolated many-body systems \cite{Nandkishore15}. An usual scenario assumes that system reveals thermalization for a long-time evolution. It means that for a small subsystem, the remaining (large) part of the system acts as a  reservoir. In such a situation the standard statistical mechanics gives an appropriate description of the long-time evolution of the system studied as follows from eigenstate thermalization hypothesis \cite{Deutsch91,Srednicki94}. However, if the system is placed in a disordered media, the described above   scenario may be simply incorrect. MBL prevents  thermalization  and the system does not reach a thermal equilibrium characterized by a loss of the memory of the initial state. Instead, as shown experimentally in cold atomic experiments \cite{Schreiber15} the system remembers its initial state. This is the reason why MBL is sometimes regarded as a phenomenon that could be potentially applied to build a 
quantum memory. 

While the existence of MBL phase for sufficiently strong quenched disorder is now well established,  the understanding of the transition between the localized and extended phases is far from being complete. The crossover between ergodic and MBL phase lays somehow beyond  the standard equilibrium statistical mechanics. It should be understood as a dynamical phase transition \cite{Vosk} between a region where the system fulfills standard statistical mechanics (and thermalizes) and a region where thermalization does not take place. Furthermore the nature of eigenstates  is also changed during this transition. The entanglement entropy for thermal states scales volume-like, but for MBL states area law scaling is postulated \cite{ImbrieReview,igloi,serbynentropy,Bardarson,Khemani2017,Huse14,ROS2015420}. 
Most  of theoretical and numerical results were obtained for Heisenberg model \cite{znidaric2008,baygan2015,Enss17,Alet15,Imbriediag,Janarek18},
typically  for small systems because of computational limitations. Time-evolution from specially prepared initial states may be studied using time-dependent Density Matrix Renormalization Group (tDMRG) for one-dimensional (1D)  systems of considerable sizes. Let us also note the possibility to treat a system in the thermodynamic limit for binary discrete  disorder \cite{Enss17}. The real limitation for time dynamics studies emerges due to the entanglement growth which limits the times reliably reached, especially close to the transition. Such studies are performed for spin systems but also for models emulating bosons
\cite{Sierant17,Sierant17b,Sierant18} or fermions \cite{Mondaini15} in a lattice. 

In this article we consider a model of the crossover between ergodic and MBL phase for small 2D systems being inspired by 1D spin studies \cite{Alet15,Mondaini15,Bertrand16,Serbyn16}. We analyze the statistics of energy levels in the transition between localized and extended phases.  Limiting ourselves to spin chains we attempt, using exact diagonalization studies, to compare systems of different geometry.  This is also inspired by recent experiments with many body localization for two-dimensional (2D) and three-dimensional cold atom systems \cite{Kondov15,Choi16,Bordia16,Bordia2DMBL} as well as theoretical discussions \cite{PhysRevB.95.155129,PhysRevB.94.144203}.

Exact diagonalization imposes a severe constraint on the size of the system we study. We shall consider a system with
16 spins in different "lattice geometries":  a standard one-dimensional chain, 2D ladder-like model  with 8 rungs (2x8 system), 2D - 4x4 square lattice and 2D - triangular case, all with periodic boundary conditions (PBC) (compare Fig.\ref{Fig1}). 
While it has been recognized that PBC may lead to additional symmetries for small systems with a discrete disorder \cite{Janarek18} following this study we are confident that for a continuous random disorder PBC are the appropriate choice that helps us to mimic larger systems. Two possible disorder realizations are discussed by us: an uniform disorder as well as the quasi-disorder, induced by the appropriate arrangement of incommensurate frequencies of laser beams involved - the model leading to the so called  Aubry-Andre\cite{aubry1980} situation for a single particle physics.  

\begin{figure*}
\includegraphics[width=2\columnwidth]{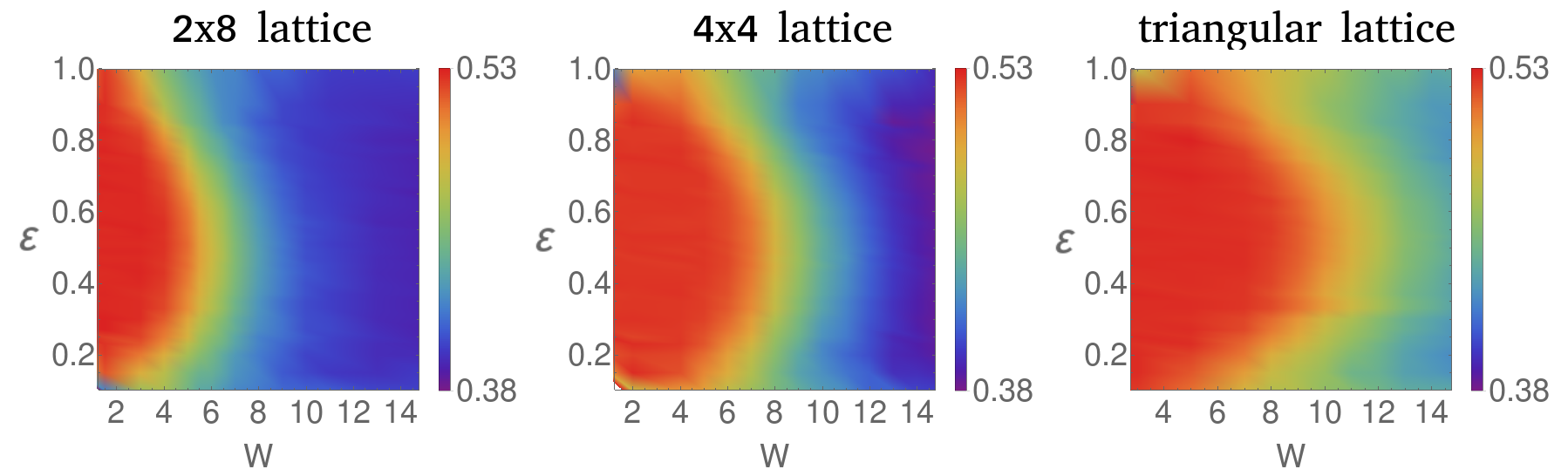}
\caption{Color maps of the averaged gap ratio $\bar r$ in function of disorder amplitude and relative energy  $\varepsilon$ \eqref{defeps}. Observe that the crossover from GOE to Poisson statistics depends on the system geometry. Results are obtained for random uniform disorder with PBC for configurations: 2x8, 4x4 and the triangular one.} 
\label{map}
\end{figure*}

\begin{figure}
\includegraphics[width=\columnwidth]{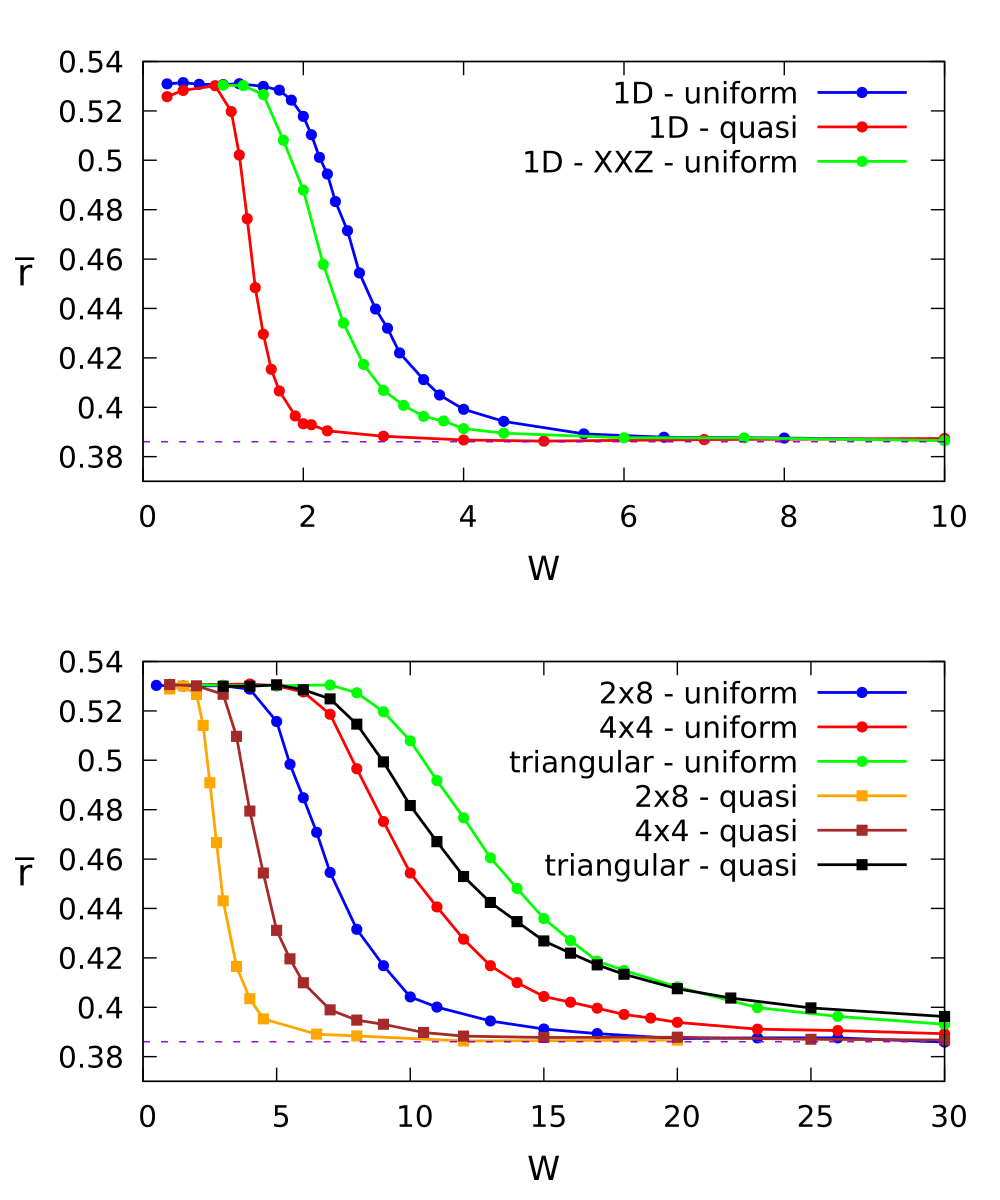}
\caption{Average gap ratio $\bar r$  \eqref{av_r} averaged over 800-1500 realizations of uniform or quasi-disorder for eigenvalues included in $\varepsilon \in (0.49,0.51)$ is presented for mentioned lattices. It shows directly that statistics of eigenvalues changes from GOE to Poisson distribution during the crossover. Top row corresponds to details of 1D model: XXX, XXZ with uniform disorder and XXX with quasi-disorder. In the bottom row there is presented comparison of averaged $\bar r$ for geometries and disorders mentioned in the article.}
\label{Fig3}
\end{figure}

\section{The model}
All calculations are performed for the same Heisenberg model which is defined by the following Hamiltonian:
\begin{equation}\label{hamilto1}
H=\sum_{<i\ne j>}[ J(S^x_i S^x_{j}+S^y_i S^y_{j})+J_z S^z_i S^z_{j} ]+ \sum_i h_i S^z_{i}
\end{equation}
where $<i\ne j>$ indicates the sum over nearest neighbors (in which each pair appears once only). 
The first part of the Hamiltonian describes interactions between neighbouring spins with a possible breaking of the spherical symmetry in the z-direction. 
We assume $J=1$ as a unit of energy taking
 $J_z=1$ for the isotropic XXX case  and $J_z=0.5$ for XXZ model.
 
The second part refers to an external field in the z-direction which is random and produces disorder in the system.
We consider two cases.  For the ``uniform disorder''   $ h_i \in [-W,W]$ is drawn  from the uniform random distribution.
We consider also the case of quasi-periodic disorder  defined in the 1D chain \cite{Iyer13,Schreiber15,Naldesi16,Khemani17a} as:
\begin{equation}
h_i=W\cos(2\pi \tau i+\phi)
\end{equation}
where: $W$ is the disorder amplitude, $i$ the site number, $\tau=\frac{1+\sqrt{5}}{2}$ and a  phase: $\phi \in [-\pi,\pi]$. Each realization corresponds to a fixed $\phi$,
averages over quasi-periodic disorder correspond to averaging with respect to $\phi$ chosen randomly and uniformly.
 
The definition of quasi-periodic disorder has to be modified for 2D  models depicted in 
Fig.\ref{Fig1}(b-d). In particular for the ladder system (with eight rows and two columns) let us define index $i$ as a pair ${j,k}$ with $j$ numbering rows and $k$ - columns.
Then we take  
\begin{equation}
h_{i}=W\cos(2\pi \tau j+\phi_k)
\end{equation}
where: $\tau=\frac{1+\sqrt{5}}{2}$, and the phases: $\phi_k$ $\in [-\pi,\pi]$ are independent for each column. The same construction holds for the square lattice model (4x4)
with four independent $\phi_k$ for each realization.

For a triangular lattice model case there is a need to define a model of  disorder which could be realized in a cold-atom experiment \cite{Sengstock}.
We assume that the quasi-random disorder in the triangular lattice is created by adding a weak incommensurate lattice:
\begin{equation}
\begin{split}
V(r)=\frac{1}{2}W[\cos(2\pi \tau {\mathbf b_3}\cdot{\mathbf r}-\phi_1)+\cos(2\pi \tau {\mathbf b_2}\cdot{\mathbf r}-\phi_2)\\
+\cos(2\pi \tau {\mathbf b_1}\cdot{\mathbf r}+\phi_1-\phi_2)]
\end{split}
\end{equation}
where direction vectors ${\mathbf b_1}=\left ( 0 , -\sqrt{3} \right )$, ${\mathbf b_2} = \left( 3/2, -\sqrt{3}/2\right)$,  ${\mathbf b_3}=\left (3/2, \sqrt{3}/2\right)$ and 
$ {\mathbf r}=\left( x,y\right)^T  $ corresponds to position of spins
with random phases: $\phi_1,\phi_2 \in [-\pi,\pi]$.

While the implementation of PBC for ladder or square models is straightforward, it is by no means so for the triangular lattice (compare Fig.~\ref{Fig1}). The same letters in Fig.~\ref{Fig1}d identify the links connected in the PBC construction.

\section{Level statistics and MBL}

\begin{figure}
	\includegraphics[width=\columnwidth]{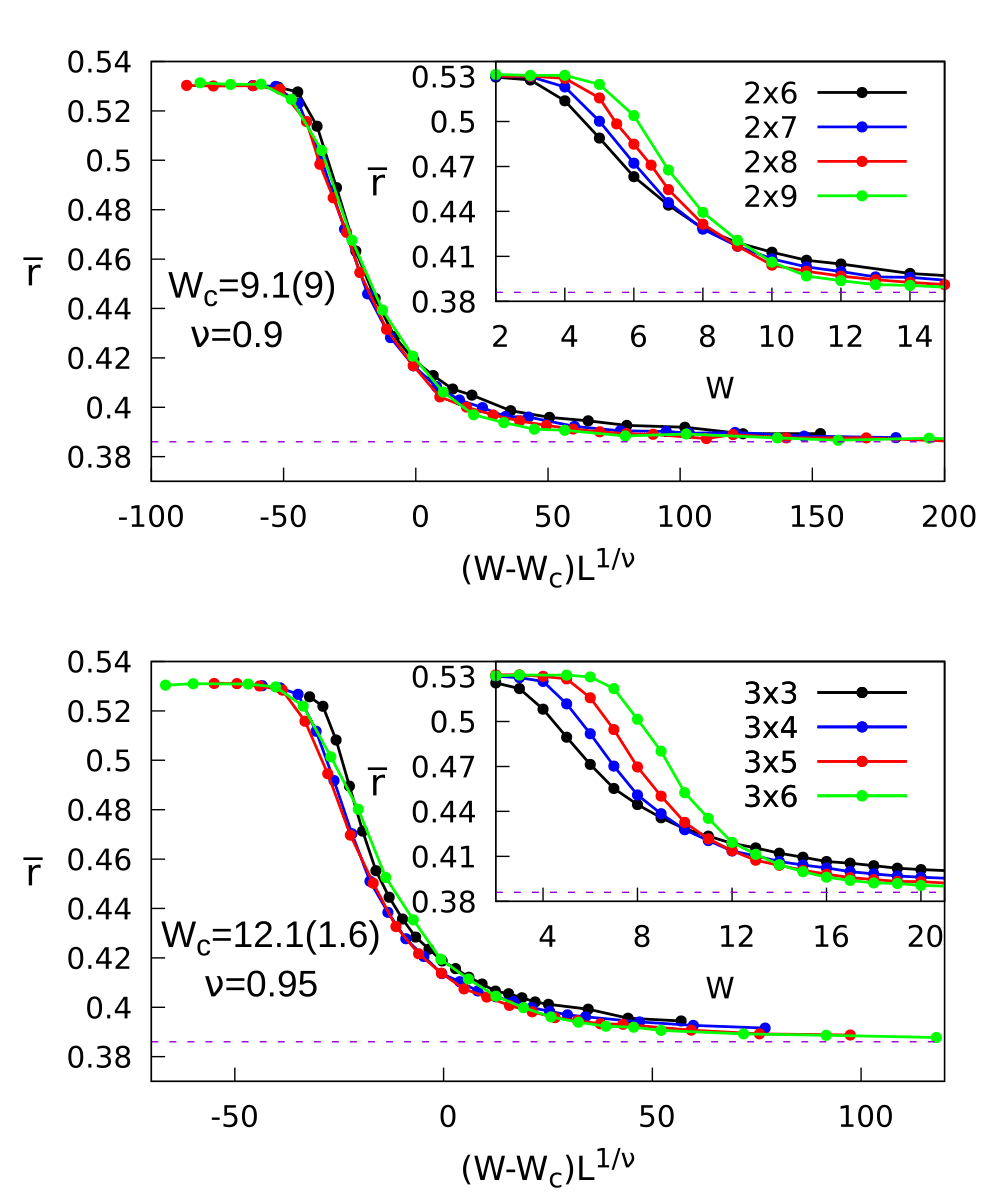}
	\caption{Finite size scaling for the average gap ratio $\bar r$  \eqref{av_r} for uniform disorder (eigenvalues within $\varepsilon \in (0.49,0.51)$ range are included; number of disorder realizations varied between 250-100000 depending on the matrix size). Upper panel shows results for ladder system, the lower panel shows 3x$L$ rectangular lattice.  Critical disorder values (indicated in the figure) $W_c$ are obtained from crossing points between curves forming original data shown in the insets.}
	\label{Fig4}
\end{figure}

To characterize the spectra in extended and MBL regime as well as in the crossover between them we use tools borrowed from quantum chaos and random matrix  theory \cite{Mehtabook,Haake}. While quantum chaos studies considered typically the so called spacing distributions i.e. the distribution of spacings between consecutive energy levels (unfolded to have the mean unit density \cite{Haake}) a simpler measure called the gap ratio distribution may be introduced \cite{Oganesyan07}. The dimensionless gap ratio is defined as  the ratio of consecutive level spacings $s_n$ corresponding to nth energy level:  
\begin{equation}
\label{av_r}
 r_n=\frac{\min(s_n,s_{n-1})}{\max(s_n,s_{n-1})}.
\end{equation}
A set of numbers $\{r_n\}$ may be thought as drawn from some distribution $P(r)$ (we drop subscript $n$ in the following writing simply $r$). In particular it was found that 
for typical ergodic situation represented by Gaussian Orthogonal Ensemble (GOE) of matrices the average $\bar r_{\mathrm GOE} \approx 0.53$ while the Poisson ensemble of uncorrelated levels (integrable limit) $\bar r_{\mathrm POI} \approx 0.39$ \cite{Oganesyan07}. Several numerical studies used the averaged gap ratio, $\bar r$, to characterize the degree of localization \cite{Alet15,Mondaini15,Naldesi16}. Much more rare were studies of the full $P(r)$ distribution.  Interestingly, it was found theoretically  that one may obtain an analytic expression for the Poisson limit
as well as an approximate formula for the GOE obtained considering smallest three by three matrices \cite {Atas13} (recall that the famous Wigner spacing distribution for GOE
is obtained considering 2x2 matrices). The corresponding expression reads for GOE:
\begin{equation}
\label{goe}
P(r)=\frac{27}{4}\frac{ r+ r^2}{(1+ r+ r^2)^{5/2}}.
\end{equation}
The approximate formula for large matrix sizes is also available \cite{Atas13}. 

In the opposite limit one may obtain an analytic result not only for a pure  Poisson but for a class of systems leading to the generalized semi-Poisson ensemble \cite{Bogomolny99}
by considering particles on a ring interacting with nearest neighbors only \cite{Atas13a}. The corresponding expression reads

\begin{equation}
\label{r}
P( r,\beta)=\frac{2\Gamma(2\beta+2)\Gamma^2(\beta+2)}{(\beta+1)^2\Gamma^4(\beta+1)} \frac{ r^{\beta}}{( r +1)^{2\beta+2}}
\end{equation}
Level repulsion parameter $\beta$ is restricted to values: $0 \leq \beta \leq 1$, $\beta=0$ corresponds to Poisson distribution while $\beta=1$ corresponds to the so called
semi-Poisson ensemble.

We mention in detail the generalized semi-Poisson case for $P(r)$ since recently an interesting study analyzing level spacing statistics  has been presented by Serbyn and Moore \cite{Serbyn16}.  They found out, that to a good approximation, the distribution
\begin{equation}
\label{plasma}
P(s,\beta,\gamma)=C_1s^{\beta}e^{-C_2 s^{2-\gamma}}
\end{equation}
(where $s$ refers to spacing between consecutive energy levels) quite well describes the spacing distribution in the transition regime between MBL
and the extended regime for the spin system studied.  The coefficients $C_i$ are determined by the normalization and the requirement that  the average $\langle s\rangle=1$. The parameters $\beta$ and $\gamma$ are restricted to $\beta,\gamma \in[0,1]$. $\beta$ measures level repulsion at small spacings. Interestingly for $\gamma=1$
the spacing distribution corresponds to the generalized Poisson regime discussed above.

Several remarks are in order. The distribution \eqref{plasma} arrives from  
 the so called plasma model \cite{Kravtsov95}. The model predicts that for large spacings their distribution falls off as an exponent to a fractional power interpolating between the power two characteristic for gaussian ensemble and power one corresponding to the Poisson family. The original model links this large spacing behavior with  an appropriate growth of the number variance, behaving like $N^\gamma$ for large number of levels, $N$. So $\gamma$ measures decorrelation of levels at large distances. 
 A postulate of small spacing behavior as $s^\beta$ comes from general understanding of random matrices.   In the limit $\beta=1$ $\gamma=0$, Eq.\eqref{plasma} reduces to a famous Wigner
 distribution for GOE as derived for $2\times 2$ matrices. This simple form is a good approximation only, the proper distribution for GOE has been derived by Mehta \cite{Mehtabook}.
 It is known that for a sufficient statistics one can easily differentiate between the Wigner formula and the proper distribution \cite{Haake}.  Thus the expression \eqref{plasma} may be at best approximate only. Moreover, recent studies \cite{Sierant18s,Sierant18l} identify ensembles which capture MBL transition to a high accuracy - but their practical application is cumbersome. On the other hand, due to its simplicity, the plasma model approximate formula \eqref{plasma} may serve as a first indicative tool in the study of spacings. We shall use it, therefore, in the following.

\begin{figure}
\includegraphics[width=\columnwidth]{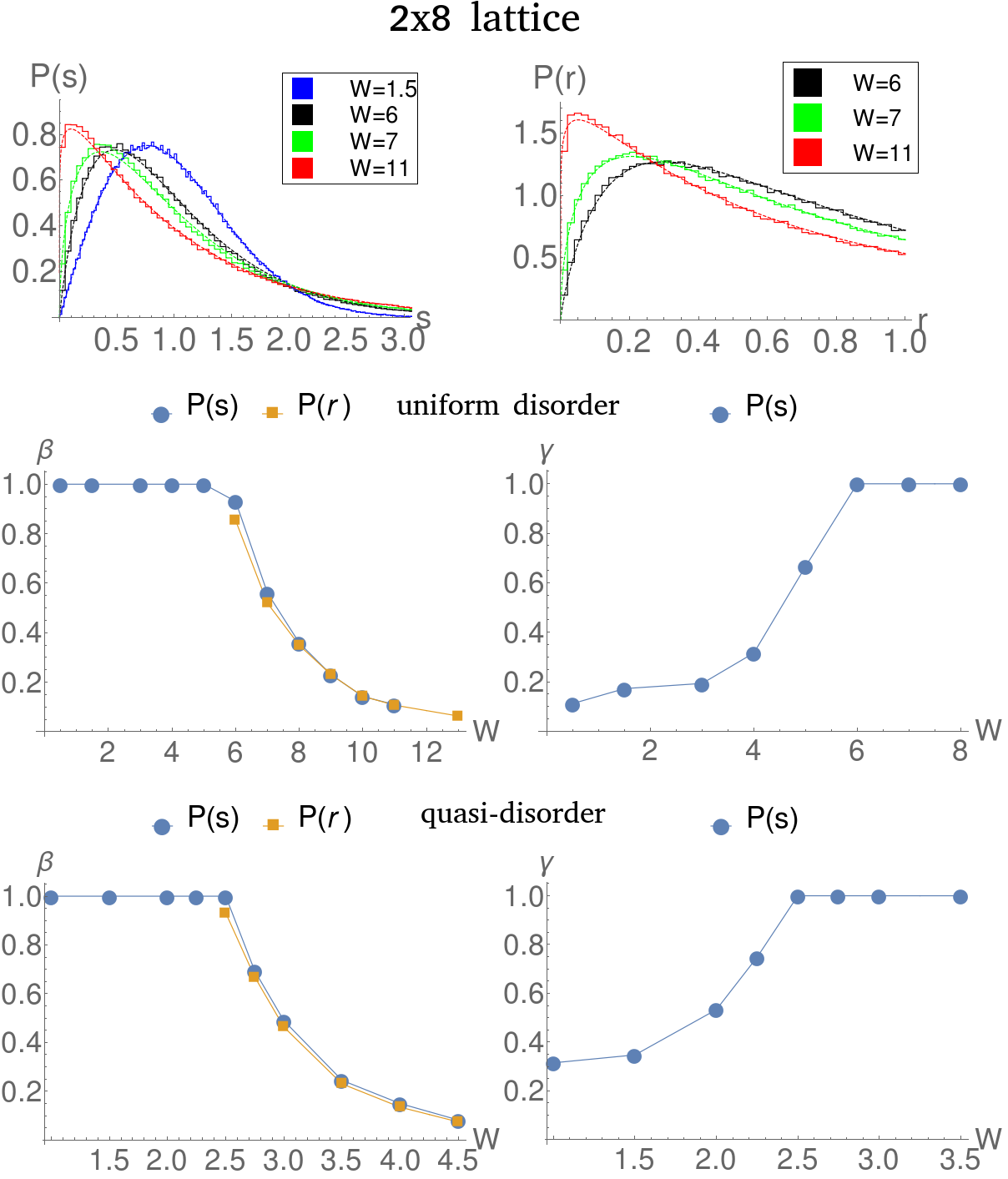}
\caption{Distributions $P(s)$ and $P(\bar r)$ for 2x8 lattice based on formulas \eqref{plasma} and \eqref{r} with fitted parameters $\beta$ and $\gamma$. Results were calculated for uniform disorder and quasi-disorder.}
\label{Fig5}
\end{figure}

\begin{figure}
\includegraphics[width=\columnwidth]{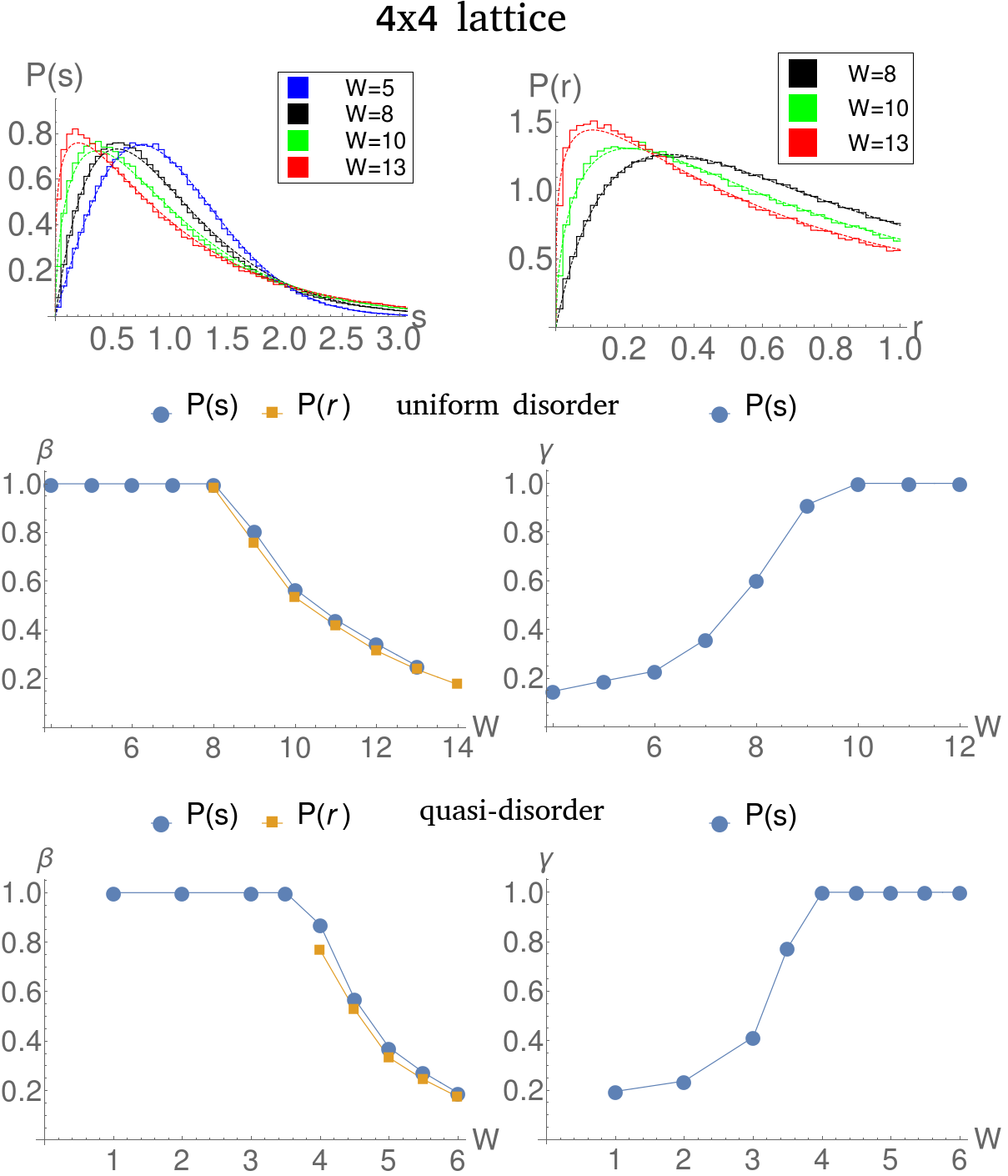}
\caption{Distributions $P(s)$ and $P(\bar r)$ for 4x4 lattice based on formulas \eqref{plasma} and \eqref{r} with fitted parameters $\beta$ and $\gamma$. Results were calculated for uniform disorder and quasi-disorder.}
\label{Fig6}
\end{figure}

\section{Results}

After presenting the systems studied and the tools used to characterize them it is a high time to discuss the results obtained. We know already that the transition region (defined by the amplitude of the disorder)  between MBL and extended states depends strongly on the energy \cite{Alet15,Naldesi16} revealing an apparent mobility edge. Being aware of this phenomenon we
rescale obtained energy eigenvalues using the formula  
\begin{equation}\label{defeps}
\varepsilon=(E-E_{max})/(E_{max}-E_{min})
\end{equation}
 where $E_{max}$ ($E_{min}$) are the highest (lowest) energy obtained from a given diagonalization. Then collecting data from different realizations of disorder we find $\bar r$ cutting the full $\varepsilon \in [0,1]$ interval into small steps. The results are
presented in  Fig.~\ref{map} for the ladder, square and {triangular} lattices and the random uniform disorder. The picture forms a kind of the phase diagram (Ergodic - MBL phase) in the disorder strength versus energy plane. Such a map was already presented for the 1D chain in Ref.~ \cite{Alet15}. Observe that the transition from the extended to localized phase shifts to larger disorder amplitudes when a transition to 2D square system is realized. The triangular lattice is the most resistive to localization requiring largest disorder amplitude. 

This fact correlated to a large extend with the number of neighbours a given lattice site is connected with via tunneling. This number is 4 for the square case and six for the  triangular lattice. And indeed the tip of the lobe for the triangular lattice occurs for $W\approx 12$ while for the square lattice for $W\approx 8$ showing an approximate proportionality of transition $W$ amplitude to the number of connected neighbors (recall that for a 1D case the transition occurs for $W\approx 3.7$ corresponding to two neighbors \cite{Alet15}).
Such a scaling indicates that a mean field analysis of MBL in dimensions bigger than one may be quite justified supporting the clams of Ref.~\cite{Yan17}.

It is clear that the behavior of the spin systems, regardless of the geometry, strongly depends on energy for small system sizes studied here. Therefore, we shall restrict  to the middle energy interval $\varepsilon \in [0.49,0.51]$ in the following. 
Fig.~\ref{Fig3} shows the dependence of the mean $\bar r$ on the strength of the disorder for different systems. The top panel presents data for a strictly 1D chain. We observe that the transition depends on the system (XXX or XXZ) studied, the transition to localized phase for the quasi-periodic disorder occurs faster than for the truly random case. 
Also this transition seems to be noticeably more rapid. In our case that is partially due to the distribution of disorder which, for the quasi-periodic case favors extremal values. However the similar behavior was observed even when both quasi-periodic and random cases were given the same
shape of the distribution \cite{Khemani17a}. 
Ref.~ \cite{Khemani17a} gives further arguments towards the claim that MBL transition for random and quasi-periodic disorder has different character. On the other hand the difference in behavior may be, at least partially, related to the presence of rare Griffiths regions for the uniform disorder 
 \cite{Lueschen17a}. 

The 2D systems are shown in the lower panel. The fact that quasi-periodic disorder leads to a much faster MBL is confirmed also for 2D situations. Again, as in Fig.~\ref{map} we may observe an important role of the number of nearest neighbors - here the triangular lattice is clearly the most resistant to localization.

Even for such small systems an attempt may be made at finite size scaling, as performed for, e.g. 1D systems \cite{Alet15,Alet17}. The average gap ratio $\bar r$  \eqref{av_r} strongly depends on the system size (compare Fig.~\ref{Fig4} where we present the results for $2\times L$ and $3\times L$ ladder models). Similarly like in \cite{Alet15} the transition becomes sharper with increasing size $L$. The finite size scaling yields both the critical disorder amplitude $W_c$ and the critical exponent $\nu$.
Interestingly, the obtained exponents are close to the value $0.91$ obtained for 1D system \cite{Alet15}.

\begin{figure}
\includegraphics[width=\columnwidth]{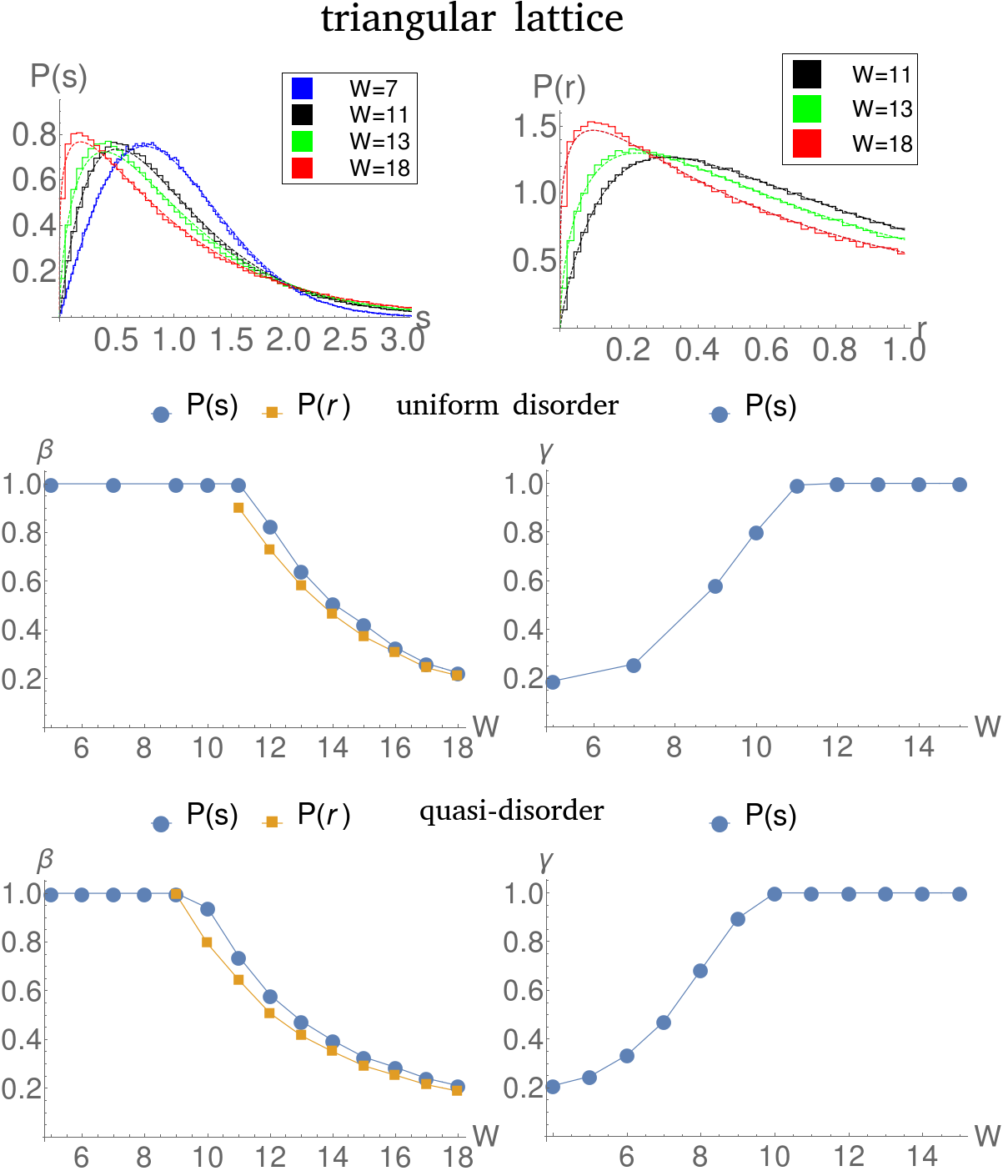}
\caption{Distributions $P(s)$ and $P(\bar r)$ for triangular lattice based on formulas \eqref{plasma} and \eqref{r} with fitted parameters $\beta$ and $\gamma$. Results were calculated for uniform disorder and quasi-disorder.}
\label{Fig7}
\end{figure}

As mentioned in the previous Section we have at our disposal not merely the $\bar r$ values for the center of the spectrum - we may consider the full gap ratio distribution $P(r)$ or the corresponding spacing distribution, $P(s)$.
Using models \eqref{plasma} and  \eqref{r} we present in (Fig.~\ref{Fig5}-\ref{Fig7}) how statistics of energy levels change in the crossover regime.
Each  figure is constructed in the similar way facilitating a comparison of systems with different shapes. The first row represents exemplary histograms fitted with the distribution of the consecutive level spacings,  \eqref{plasma} (left) and the gap ratio, \eqref{r}  (right)  for the uniform random disorder. The second row shows the dependence of the fitted level repulsion parameter  $\beta$ and the level decorrelation parameter $\gamma$ (compare \eqref{plasma}) on the disorder amplitude for the uniform random disorder. The third row presents  the same parameters but for the quasi-random disorder case. 

Results shown in Fig.~\ref{Fig5}-\ref{Fig7} indicate that the scenario of the crossover has a similar character independently of the shape of the model studied. Generally, as in Fig.~\ref{map}, we observe the shift of the crossover regime towards larger disorder amplitudes with increasing number of neighbors in different models. But also, interestingly, one may roughly determine two distinct regimes.  Close to the fully developed MBL
phase $\gamma=1$ indicating exponential tail of the spacings while $\beta$ smoothly grows from its Poisson distribution value $\beta=0$ in the MBL phase up to $\beta=1$.
Once the full available value (for time-reversal invariant systems) of short range level repulsion $\beta=1$ is reached only then levels start to correlate over large distances, and for $\beta=1$ 
level decorrelation parameter $\gamma$ decreases gradually to zero. That corresponds to the change of the tail for large spacings from the exponential to gaussian-like, characteristic for GOE.
 Depending on the model, there  may be a small interval of disorder values where both $\beta$ and $\gamma$ changes simultaneously close to $\beta,\gamma=1$ - the precise
 determination of this region would require much better statistics. Since anyway our data are indicative only as the transition region is known to change with the system size (see e.g. \cite{Alet15,Khemani17a}) we do not further investigate this point. Let us just mention that a similar behavior has been observed for interacting bosons \cite{Sierant17b,Sierant18}. 

\begin{figure}
\includegraphics[width=\columnwidth]{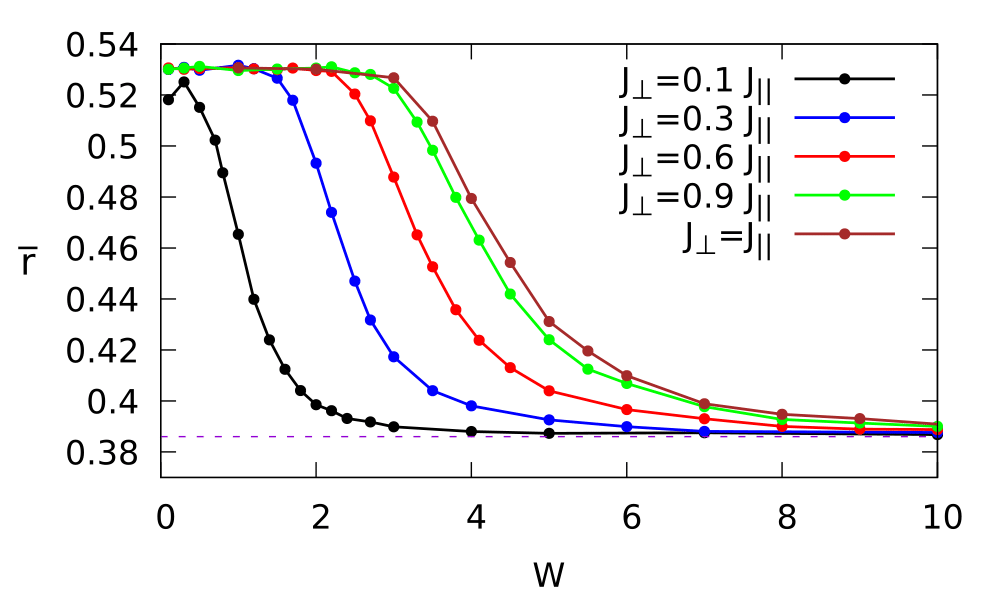}
\caption{Average gap ratio $\bar r$  \eqref{av_r} averaged over 800-1500 realizations of quasi-disorder for eigenvalues included in $\varepsilon \in (0.49,0.51)$. The localized to extended states transitions shifts gradually from 1D to 2D value when the vertical coupling between horizontal 1D chains is increased.  }
\label{Fig8}
\end{figure}

Let us also observe  in the ``close to MBL'' region of $\gamma=1$  the predictions of the generalized semi-Poisson ensemble consistently well describe both the spacings and the gap ratio distribution as shown by blue and yellow dots representing the fits. In the second regime, close to the extended, GOE-like phase, where $\beta=1$ and $\gamma$ is changing no analytic formula is known for the gap ration distribution, $P(r)$. Thus in this regime only $P(s)$ is fitted.

\section{Transition from 1D to 2D}

In cold atomic systems it is quite easy to control parameters of the Hamiltonian, for example the tunnelings in different directions. In the recent experiment \cite{Bordia16} several 1D chains of fermions were coupled perpendicularly modifying the 1D geometry towards two dimensions. Strong dependence of the localization on that coupling was observed. 

Such a situation may be simply realized in our model. Consider the modified XXX Hamiltonian, Eq.~\eqref{hamilto1}
\begin{equation}\label{hamilto2}
H=\sum_{i,j}[ J_{\parallel} {\vec S}_{i,j} {\vec S}_{i+1,j}+J_{\perp}{\vec S}_{i,j} {\vec S}_{i,j+1}) ]
+ \sum_{i,j} h_{i,j}S^z_{i,j} \end{equation}
in a $4\times 4$ geometry.
Here we shall consider the quasi-random disorder only with $h_{i,j} = W\cos(2\pi \tau i+\phi_j)$. For a single realization of disorder we take four different random phases $\phi_j$.
The results are averaged over several realizations of the choice of phases and are presented in Fig.~\ref{Fig8}. With an increasing coupling between columns of the system we observe a smooth transition from the case of the uncoupled 1D chains to the fully connected $4\times 4$ plaquette.

\section{Conclusions}\label{Concl}
We have analyzed the transition between extended and localized states in the standard spin model using exact diagonalizations of small systems. By different arrangement of spins
we were able to consider different geometries: a 1D chain, a ladder system, a toy model of the square or triangular lattice. Both uniform and quasi-periodic disorder were considered.

We have confirmed that for all the systems studied the transition between localized and extended states is sharper for quasi-random disorder. For a given disorder it is the number of neighbors that play the decisive role in deciding how large the amplitude of the disorder is needed to observe the transition. This points out that studies of MBL in two or three dimensions using the mean field approach are justified. By adjusting vertical and horizontal tunneling ratio one may realize a smooth transition between 1D and 2D systems. The localization border shifts then to large field amplitudes, as expected.

Finally let us mention that very small systems, amenable for exact diagonalization may be of interest on their own - experimental study of such small but extremely well controlled models when atoms may be added one by one becomes a fashionable field of its own \cite{Wenz14,Murmann15,Blume15,Barredo17,Weber18}. Adding well controlled disorder to such systems is entirely feasible.

\begin{acknowledgments} 
Discussions with Piotr Sierant were enlightening.
 We acknowledge support of the National Science Centre, Poland via project  No.2015/19/B/ST2/01028 (DW) and 2016/21/B/ST2/01086 (JZ). This work was performed also within  Horizon2020 FET project QUIC (nr.~641122).
\end{acknowledgments}

\input{mbl06.bbl}
\end{document}

%% file: mbl06.bbl
%